\documentclass{sig-alternate}

\usepackage[vlined,ruled,boxed,linesnumbered]{algorithm2e}

\usepackage{url}
\usepackage{paralist}
\usepackage{enumitem}

\SetKw{by}{by}
\SetKw{downto}{down to}
\SetKwRepeat{DoWhile}{do}{while}

\newtheorem{Thm}{Theorem}
\newtheorem{Cor}[Thm]{Corollary}
\newtheorem{Lem}[Thm]{Lemma}

\newtheorem{Rem}[Thm]{Remark}
\newtheorem{Exam}[Thm]{Example}

\newcommand{\DFT}{\mathrm{DFT}}

\newcommand{\rev}{\mathrm{rev}}
\renewcommand{\th}{\text{-th}}

\newcommand{\quo}{\ \mathrm{quo}\ }

\renewcommand{\a}{\textbf{a}}

\def\a{\textbf{a}}

\def\TFT{\mathrm{TFT}}
\def\rev{\mathrm{rev}}
\def\DFT{\mathrm{DFT}}
\def\IDFT{\mathrm{IDFT}}

\begin{document}

\conferenceinfo{ISSAC}{'13 Boston, Massachusetts USA}

\title{A new Truncated Fourier Transform algorithm\titlenote{}}

\numberofauthors{1}
\author{
\alignauthor Andrew Arnold\\
       \affaddr{Symbolic Computation Group}\\
       \affaddr{University of Waterloo}\\
       \email{a.arnold55@gmail.com}\\
       \url{http://cs.uwaterloo.ca/~a4arnold}
}

\maketitle
\begin{abstract}
Truncated Fourier Transforms (TFTs), first introduced by Van der Hoeven, refer to a family of algorithms that attempt to smooth ``jumps'' in complexity exhibited by FFT algorithms.  We present an in-place TFT whose time complexity, measured in terms of ring operations, is comparable to existing not-in-place TFT methods.  We also describe a transformation that maps between two families of TFT algorithms that use different sets of evaluation points.
\end{abstract}

\section{Introduction}

Let $\mathcal{R}$ be a ring containing an $N$-th principal root of unity $\omega$.  Given two polynomials $f,g \in \mathcal{R}[z]$, $\deg(fg)<N$, we can compute $fg$ by way of the Discrete Fourier Transform (DFT): a linear, invertible map which evaluates a given polynomial at the powers of $\omega$.

Computing the DFT naively is quadratic-time.  If, however, $N$ is comprised of strictly small prime factors, one can compute a DFT using $\mathcal{O}(N \log N)$ arithmetic operations by way of the Fast Fourier Transform (FFT).  The most widely-used FFT, the {\em radix-2} FFT, requires that $N$ is a power of two.  To compute the DFT of an input of arbitrary size, one typically appends zeroes to the input to give it power-of-two length, then applies a radix-2 FFT.  This results in significant jumps in the time and space complexities of the radix-2 FFT.

Truncated Fourier Transforms (TFTs) smooth these jumps in complexity.  A TFT takes a length-$n$ input, $n\leq N$, and returns a size-$n$ subset of its length-$N$ DFT, with time complexity that grows comparatively smoothly with $n \log n$.  Typically one chooses the first $n$ entries of the DFT, with the DFT sorted in bit-reversed order.  This is natural choice as it comprised the first $n$ entries of the output of an in-place FFT, if no re-sorting is performed.  We will call such a TFT the {\em bit-reversed} TFT.

Van der Hoeven \cite{vdH:tft} showed how one could obtain a polynomial $f(z)$ from its bit-reversed TFT, provided one knows the terms of $f(z)$ with degree at least $n$.  This allows for faster FFT-based polynomial multiplication, particularly for products whose degree is a power of two or slightly larger.  Harvey and Roche showed further in \cite{HR} how the bit-reversed TFT transform can be made in-place, at the cost of a constant factor additional ring multiplications.

Mateer \cite{Todd} devised a TFT algorithm based on a series of modular reductions, that acts as a preprocessor to the FFT.  Mateer's TFT algorithm, which we discuss in section \ref{sec:cyclotomicTFT} breaks an input $f(z)$ with $\deg(f)<n$, into its images modulo cyclotomic polynomials of the form $z^{k}+1$, $k$ a power-of-two.  We will call this TFT the {\em cyclotomic} TFT.  In \cite{Igor}, Sergeev showed how the cyclotomic TFT can be made in-place, with time complexity comparable to not-in-place TFT algorithms.  In section \ref{sec:newTFT}, we restate Sergeev's algorithm.  In section \ref{sec:ctft}, we present a new in-place algorithm, related to Sergeev's, for computing the cyclotomic TFT.

One caveat of the cyclotomic TFT is that different-sized inputs may use entirely different sets of evaluation points.  This is problematic in applications to multivariate polynomial multiplication.  In section \ref{sec:affine}, we show how an algorithm that computes the cyclotomic TFT can be modified to compute a bit-reversed TFT by way of an affine transformation.

As a proof of concept we implemented the algorithms introduced in this paper in python.  These implementations can be found at the author's website at:
\begin{center}
\url{http://cs.uwaterloo.ca/~a4arnold/tft}.
\end{center}

\section{Preliminaries}\label{sec:prelim}

\subsection{The Discrete Fourier Transform}

The {\em Discrete Fourier Transform} (DFT) of a polynomial $f(z)$ is its vector of evaluations at the distinct powers of a root of unity.  Specifically, if $f(z) = \sum_{i=0}^{N-1}a_iz^i$ is a polynomial over a ring $\mathcal{R}$ containing an $N\th$ primitive root of unity $\omega$, then we define the discrete Fourier transform of $f(z)$ as
\begin{equation}
\DFT_\omega( f ) = \left( f(\omega^j) \right)_{0 \leq j < N }.
\end{equation} 
We treat the polynomial $f$ and its vector of coefficients $a  = ( a_0, a_1, \dots, a_{N-1} )$ as equivalent and use the notation $\DFT_\omega( a )$ and $\DFT_\omega( f )$ interchangeably.  If we take addition and multiplication to be pointwise in $\mathcal{R}^N$, then the map $\DFT_\omega:\mathcal{R}[z]/(z^N-1)\rightarrow \mathcal{R}^N$ forms a ring homomorphism.  If $\omega$ is a {\em principal} root of unity, that is, for $j$ not divisible by $N$, $\sum_{i=0}^{N-1} \omega^{ij} = 0$,
then $\DFT_\omega$ has an inverse map $\IDFT_{\omega} : \mathcal{R}^N \rightarrow \mathcal{R}[z]/(z^N-1)$, defined by
\begin{equation}
\IDFT_{\omega}( \hat{a} ) = \tfrac{1}{N}\DFT_{\omega^{-1}}( \hat{a} ),
\end{equation}
where $\hat{a} \in \mathcal{R}^N$ and we again we treat a polynomial as equivalent to its vector of coefficients.  This suggests a multiplication algorithm for $f,g \in \mathcal{R}[z]$.
\begin{Thm}[The Convolution Theorem]
Suppose $\mathcal{R}$ is a ring containing an $N\th$ principal root of unity $\omega$ and let $f,g \in \mathcal{R}[z]$.  Then
\begin{equation}
fg \bmod (z^N-1) = \IDFT_\omega\left( \DFT_\omega(f)\cdot \DFT_\omega(g) \right),
\end{equation}
where ``$\cdot$'' is the vector component-wise product, and, given two polynomial $g(z), h(z) \in \mathcal{R}[z]$, $g(z) \bmod h(z)$ denotes the unique polynomial $r(z)$ such that $h(z)$ divides $g(z)-r(z)$ and $\deg(r)<\deg(h)$ throughout.
\end{Thm}
Thus to multiply $f$ and $g$, we can choose $N>\deg(fg)$ and an $N\th$ principal root of unity $\omega \in \mathcal{R}$, compute the length-$N$ DFTs of $f$ and $g$, take their pointwise product, and take the inverse DFT of the pointwise product.  
\subsection{The Fast Fourier Transform}

We can compute the Discrete Fourier †ransform of $f(z)$, $f$ reduced modulo $(z^N-1)$, by way of a {\em Fast Fourier Transform} (FFT).  The FFT is believed to have been first discovered by Gauss, but did not become well-known until it was famously rediscovered by Cooley and Tukey \cite{CT}.  For a detailed history of the FFT we refer the reader to \cite{history}.

The simplest and most widely-used FFT, the {\em radix-2} FFT, assumes $N=2^p$ for some $p \in \mathbb{Z}_{\geq 0}$.  We describe the radix-2 FFT in terms of modular reductions.   Let $N$ be a power of two.  We break $f$ into images modulo polynomials of decreasing degree until we have the images $f \bmod (z-\omega^i) = f(\omega^i)$, $0 \leq i < N$.  At the start of the first iteration we have $f$ reduced modulo $z^N-1$.  After $i$ iterations, we will have the $2^i$ images
\begin{equation}
f \bmod (z^{u}-\omega^{uj}), \hspace{0.4cm}\text{ for } 0 \leq j < 2^i,
\end{equation}
where $u=N/2^i$.  If $i=p$ this gives us the DFT of $f$.  Consider then the image $f' = f(z) \bmod (z^{2u} - \omega^{2uj}) = \sum_{i=0}^{2u-1}b_i$, for some $j$, $0 \leq j < \tfrac{N}{2u}$.  We break this image into two images $f_0$ and $f_1$, where
\begin{align}
f_0 &= f' \bmod (z^{u} - \omega^{uj}), \text{ and} \\
f_1 &= f' \bmod (z^{u} + \omega^{uj}) = f' \bmod (z^u - \omega^{uj + N/2}),
\end{align}
If we write $f' = \sum_{k=0}^{u-1}b_kz^k$, then we can write $f_0$ and $f_1$ in terms of the coefficients $b_k$:
\begin{equation}
f_0 = \sum_{k=0}^{u-1}(b_k + \omega^{uj}b_{k+u} )z^k, \hspace{0.4cm} f_1 = \sum_{k=0}^{u-1}(b_k - \omega^{uj}b_{k+u})z^k.
\end{equation}
thus, given an array containing the coefficients $b_k$ of $f'$, we can write $f_0$ and $f_1$ in place of $f'$ by way of operations
\begin{equation}\label{eqn:rad2fft}
\left[\begin{matrix} b_k \\ b_{k+u}\end{matrix}\right] \longleftarrow \left[\begin{matrix} 1 & \omega^{uj} \\ 1 & -\omega^{uj}\end{matrix}\right]\left[\begin{matrix} b_{k} \\ b_{k+u}\end{matrix}\right], \hspace{0.4cm} 0 \leq i \leq u.
\end{equation}

The pair of assignments \eqref{eqn:rad2fft} are known as a {\em butterfly operation}, and can be performed with a ring multiplication by the {\em twiddle factor} $\omega^{uj}$, and two additions.  Note $f_0$ and $f_1$ are in a similar form as $f'$, and if $u>1$ we can break those images into smaller images in the same fashion.  Starting this method with input $f \bmod (z^N-1)$, will give us $f \bmod (z-\omega^j) = f(\omega^j)$, for $0 \leq j < N$.

If the butterfly operations are performed in place, the resulting evaluations $f(\omega^j)$ will be written in {\em bit-reversed} order.  More precisely, if we let $[j]_p$ denote the integer resulting from reversing the first $p$ bits of $j$, $0 \leq j <2^p$, we have that $f(\omega^j)$ will be written in place of $a_{k}$, where $k=[j]_p$ and $\log$ is taken to be base-2 throughout.  As an example,
\begin{equation}
[11]_5 = [{\tt 01011}_2]_5 = {\tt 11010}_2 = 16+8+2=26.
\end{equation}

We can make the FFT entirely in-place by computing the powers of $\omega^u$ sequentially at every iteration.  This entails traversing the array in a non-sequential order.  Procedure {\tt\ref{proc:FFT}} describes such an implementation.

If we observe that
\begin{equation}\label{eqn:invButterfly}
\left[\begin{matrix} 1 & \omega^{uj} \\ 1 & -\omega^{uj}\end{matrix}\right]^{-1} = \frac{1}{2}\left[\begin{matrix} 1 &  1\\ \omega^{-uj} & -\omega^{-uj}\end{matrix}\right],
\end{equation}
then we can implement an inverse FFT by inverting the butterfly operations in reversed order.  We can, moreover, delay multiplications by powers of $\tfrac{1}{2}$ until the end of the inverse FFT computation.  This entails multiplying each coefficient by $\tfrac{1}{N}$.

\begin{procedure}
	\caption{FFT(\textbf{a}, $N$, $\omega$), an in-place implementation of the radix-2 Cooley-Tukey FFT}\label{proc:FFT}
	\KwIn{
		\begin{itemize}[noitemsep,nolistsep]
			\item  $\textbf{a} = (\textbf{a}_0, \dots, \textbf{a}_{N-1})$, an array vector $a \in\mathcal{R}^N$.
			\item  $\omega$, a root of $z^N-1$, with $N=2^p$
		\end{itemize}
	}
	\KwResult{ $\DFT_\omega(a)$ is written to the array ${\bf a}$ in bit-reversed order.  }\vspace{0.3cm}
	
	\For{$i \longleftarrow 1$ \KwTo $p$}{
		$u \longleftarrow N/2^i$\\
		\For{$j \longleftarrow 0$ \KwTo $\tfrac{N}{2u}-1$}{
			$t \longleftarrow 2\rev_i(j)u$\\
			\For{$k \longleftarrow 0$ \KwTo $u-1$ }{
				\tcp{Butterfly operations \eqref{eqn:rad2fft}} 
				$\left[\begin{matrix} {\bf a}_{t+k} \\ {\bf a}_{t+k+u}\end{matrix}\right] \longleftarrow \left[\begin{matrix} 1 & \omega^{uj} \\ 1 & -\omega^{uj}\end{matrix}\right]\left[\begin{matrix} {\bf a}_{t+k} \\ {\bf a}_{t+k+u}\end{matrix}\right]$ \\
				\label{line:butterfly}
			}
		}
	}
\end{procedure}

\begin{Thm}
Let $N$ be a power of two and $\omega$ an $N\th$ root of unity.  Then $\DFT_{\omega}(f)$ can be computed in place using $\tfrac{1}{2}N\log N + \mathcal{O}(N)$ ring multiplications.  The inverse FFT can interpolate $f(z)$ from its DFT with similar complexity.
\end{Thm}

Using the radix-2 FFT, if $d = \deg(fg)$, we choose $N$ to be the least power of $2$ exceeding $d$.  This entails appending zeros to the input arrays containing the coefficients of $f$ and $g$ respectively.  By this method, computing a product of degree $2^p$ costs roughly double that a product of degree $2^p-1$.  Crandall's {\em devil's convolution} algorithm \cite{scicomp} somewhat flattens these jumps in complexity, though not entirely.  It works by reducing a discrete convolution of arbitrary length into more easily computable convolutions.  More recently, truncated Fourier transform (TFT) algorithms, described hereafter, have addressed this issue.
\subsection{Truncated Fourier Transforms}

In many applications, is it useful to compute a {\em pruned} DFT, a subset of a length-$N$ DFT, at a cost less than that computing a complete DFT.  In 2004, van der Hoeven showed in \cite{vdH:issac04} that, given some knowledge of the form of the input, one can invert some pruned DFTs.  The inverse transform relies on the observation that, given any two of the inputs/outputs to a butterfly operation \eqref{eqn:rad2fft}, one can compute the other two values.  Suppose $n \in \mathbb{Z}$ is arbitrary and $N$ is the least power of two at least $n$.  For $\omega$, a primitive root of unity of order $N=2^p$, van der Hoeven showed how to invert the length-$n$ bit-reversed Truncated Fourier Transform,
\begin{equation}
\TFT_{\omega, n}( f ) = \left( f( \omega^{[i]_p} ) \right)_{0 \leq i < n},
\end{equation}
when we know the coefficients of $f(z)$ of degree at least $n$ (e.g. when $\deg(f)<n$ ).  To distinguish this particular TFT we will call it the {\em bit-reversed} TFT.
\begin{Thm}[Van der Hoeven, \cite{vdH:tft}]\label{thm:vdh}
Let $N=2^p$, $n \in (N/2,N]$, $f(z) \in \mathcal{R}[z]$ and let $\omega \in \mathcal{R}$ be a principal $N$-th root of unity.  Suppose $\deg(f)<n$.  Then $\TFT_{\omega,n}(f)$ can be computed using $n \log n + \mathcal{O}(n)$ ring additions and $\tfrac{1}{2}n\log n + \mathcal{O}(n)$ multiplications by powers of $\omega$.

$f(z)$ can be recovered from $\TFT_{\omega,n}(f)$ using $n \log n + \mathcal{O}(n)$ {\em shifted} ring additions and $\tfrac{1}{2}n\log n + \mathcal{O}(n)$ multiplications.
\end{Thm}
A shifted ring addition merely means in this context an addition plus a multiplication by $\tfrac{1}{2}$.  Van der Hoeven's algorithm generalizes to allow us to compute arbitrary subsets of the DFT.  Given a subset $S \subseteq \{0,1,\dots, N-1\}$, we define
\begin{equation}
\TFT_{\omega, S}( f ) = \left( f( \omega^{[i]_p} )\right)_{i \in S},
\end{equation}
where $f$ we now assume $f$ is of the form $f(z) = \sum_{i \in S}a_iz^i$.  However, such a transform may have a greater complexity than stated in theorem \ref{thm:vdh}, taking $n = \#S$.  Moreover, such a map is not necessarily invertible.
\begin{Exam}
Taking $\omega$ an $N=8\th$ principal root of unity with $S = \{0,3,4,5\}$ gives an uninvertible map.  To see that the map $\TFT_{\omega,S}$ is not invertible, one can check that the polynomial $f(z) = (1+\omega^2)z^5 - z^4  + (1-\omega^2)z^3 - 1$ evaluates to $0$ for $z=\omega^k$, $k \in \{ [0]_3, [3]_3, [4]_3, [5]_3\} = \{0,6,1,5\}$.
\end{Exam}
Van der Hoeven's method still exhibits significant jumps in space complexity, as it space for $N$ ring elements regardless of $n$.  In 2010, Harvey and Roche \cite{HR} introduced an in-place TFT algorithm, requiring $n + \mathcal{O}(1)$ ring elements plus an additional $\mathcal{O}(1)$ bounded-precision integers to compute $\TFT_{\omega, n}(f)$.  Their method potentially requires evaluating polynomials using linear-time methods.  This results in a constant factor additionally many ring multiplications in the worst-case.

\begin{Thm}[Roche, Theorem 3.5 of \cite{Rthesis}]
Let $N,n,f$ and $\omega$ be as in theorem \ref{thm:vdh}.  Then $\TFT_{\omega,n}(f)$ can be computed in-place using at most $\tfrac{5}{6}n \log n + \mathcal{O}(n)$ ring multiplications and $\mathcal{O}(n \log n)$ ring additions.
\end{Thm}
The inverse in-place transform entails similarly many ring multiplications and $\mathcal{O}(n \log n)$ shifted ring additions.  As an application, Harvey and Roche used this transform towards asymptotically fast in-place polynomial multiplication.

\section{The cyclotomic TFT}\label{sec:cyclotomicTFT}

\subsection{Notation}

We use the following notation throughout section \ref{sec:cyclotomicTFT} and thereafter.  Suppose now that we have a polynomial $f(z) = \sum_{j=0}^{n-1}a_jz^j \in\mathcal{R}[z]$ of degree at most $n-1$, where $n$ can be any positive integer.  For the remainder of this paper, we write $n$ as 
$n = \sum_{i=1}^{s}n_i$, where $n_i = 2^{n(i)}$, $n(i) \in \mathbb{Z}_{\geq 0}$ and $n_i > n_j$ for $1 \leq i < j \leq s$.  Again $N$ is the least power of two at least $n$.  For $1 \leq i \leq s$, we let
\begin{equation}
\Phi_i =  z^{n_i}+1.
\end{equation}

The TFT algorithms of sections \ref{sec:cyclotomicTFT} and thereafter will compute the evaluations of $f(z)$ at the roots of $\Phi_i$.  Namely, if we fix a canonical root $\omega=\omega_1$ of $\Phi_1$, and then let, for $2 \leq i \leq s$, $\omega_i=\omega_1^{n(1)/n(s)}$, a root of $\Phi_i$, these algorithms will compute
\begin{equation}
f(\omega_i^{2j+1})  \hspace{0.5cm}\text{for}\hspace{0.5cm} 0 \leq j < n_i, \hspace{0.5cm} 1 \leq i \leq s.
\end{equation}
It is the evaluation of $f(z)$ at the roots of the cyclotomic polynomials $\Phi_i$.  As such, we will call it here the {\em cyclotomic} TFT and denote it by $\TFT'_{\omega,n}(f)$.  The choice of $\omega$ only affects the ordering of the elements of the cyclotomic TFT.  If we let
\begin{equation}
S(n) = \{ n_i \leq k < 2n_i : 1 \leq i \leq s \},
\end{equation}
then we have that $\TFT'_{\omega,n}$ uses the same set of evaluation points as $\TFT_{\omega, S(n)}(f)$.

The roots of unity used as evaluation points in $\TFT'_{\omega,n}(f)$ will differ from those used in $\TFT_{\omega,n}(f)$.  For instance, $\TFT_{\omega,n}(f)$ will always use all roots-of-unity of order dividing $n_1$, whereas $\TFT'_{\omega,n}(f)$ will instead use all roots of unity of order $2n_1$.  We define, for $1 \leq i \leq s$, the images
\begin{equation}
f_i = f \bmod \Phi_i.
\end{equation}

The algorithms for computing a cyclotomic TFT all follow a similar template: we will produce the images $f_i$ sequentially, and then write the evaluation of $f$ at the roots of $\Phi_i$ in place of each image $f_i$.
\subsection{Discrete Weighted Transforms}

Given the images $f_i = f \bmod \Phi_i$, one can evaluate $f$ at the roots of $\Phi_i$ by way of a {\em Discrete Weighted Transform} (DWT), which comprises an affine transformation followed by an FFT \cite{DWT}.

In a more general setting, suppose we have an image $f^* = f(z) \bmod (z^N-c)$, where $N$ is a power-of-two.  Assuming $c$ has an $N\th$ root over our ring $\mathcal{R}$, the roots of $f(z)$ are all of the form $c^{1/N}\gamma^i$, $0 \leq i < N$, where $\gamma$ is an $N\th$ primitive root of unity.  Thus to evaluate $f$ at the roots of $(z^N-c)$ one can replace $f^*(z)$ with $f^*(c^{1/N}z)$, and then compute $\DFT_{\gamma}( f^*(c^{1/N}z))$.  Replacing $f^*(z)$ with $f^*(c^{1/N}z)$ merely entails multiplying the degree-$k$ term of $f^*$ by $c^{k/N}$, for $0 \leq k < N$.  This, combined with computing the powers of $c^{1/N}$ sequentially, entails fewer than $2N$ ring multiplications.

Thus, to evaluate $f$ at the roots of $\Phi_i(z) = z^{n_i}-\omega_i^{n_i}$, one would write $f_i( \omega_i z )$ in place of $f_i(z)$, then compute $\DFT_{\omega_i^2}( f_i( \omega_i z) )$ by way of the FFT.  As both the FFT and the affine transformation are invertible, a Discrete Weighted Transform is easily invertible as well.

\begin{procedure}
	\caption{DWT(\textbf{a}, $N$, $\omega$, $v$), the Discrete Weighted Transform}\label{proc:DWT}
	\KwIn{
		\begin{itemize}[noitemsep,nolistsep]
			\item  $\textbf{a} = (\textbf{a}_0, \dots, \textbf{a}_{N-1})$, an array containing the coefficients of $f(z)=\sum_{i=0}^{N-1}a_iz^i \in \mathcal{R}[z]$.
			\item  $\omega \in \mathcal{R}$, an $N\th$ principal root of unity.
			\item  $v \in \mathcal{R}$, a weight.
		\end{itemize}
	}
	\KwResult{
		$\TFT'_{\omega,n}(f)$ is written in place of $f$
	}
	
	\lFor{$i \leftarrow 0$ \KwTo $N-1$ }{ $\a_i \longleftarrow v^i\a_i$ }\\
	\ref{proc:FFT}($\a, N, \omega$)
\end{procedure}
\subsection{Mateer's TFT algorithm}

Algorithm \ref{alg:mat} gives a short description of Mateer's algorithm \cite{Todd} for computing $\TFT_{\omega,n}'(f)$.

\begin{algorithm}
\caption{Mateer's algorithm for computing $\TFT_{\omega,n}'(f)$}\label{alg:mat}
\KwIn{
	$n<N$, $f(z) \in \mathcal{R}[z]$, $f$ reduced modulo $z^N-1$
}
\KwResult{ $\TFT'_{\omega, n}(f)$ is written in place of $f$ }
	$(i,K) \longleftarrow (1,N)$\\

	\While{$K > n_s$}{
		Separate $f \bmod (z^K-1)$ into the images $f \bmod (z^{K/2}-1)$ and $f \bmod (z^{K/2}+1)$.\\
		\If{ $K/2 = n_i$}{
			Note $\Phi_i = z^{K/2}+1$.  Use a DWT to write the evaluation of $f$ at the roots of $\Phi_i$ in place of $f_i = f \bmod \Phi_i$.\\
			$i \longleftarrow i+1$
		}
		$K \longleftarrow K/2$
	}

\end{algorithm}

On input we are given $f(z)$ reduced modulo $z^N-1$.  If $N=n=n_1$ we can just perform a DWT on $f$ to get $\TFT_{n,\omega}'(f)$, so suppose $n$ is not a power of two and $n<N$.  We reduce the image into two images $f(z) \bmod (z^{N/2}-1)$ and $f(z) \bmod (z^{N/2}+1)$, in the same manner as described in the FFT.  This process strictly entails ring additions.  If $f(z)=\sum_{i=0}^{N-1}a_iz^i$, then $f_0$ and $f_1$ can be computed by way of $N/2$ butterfly operations.
\begin{align}
f_0(z) &= \sum_{i=0}^{N/2}(a_i + a_{i+N/2})z^i, \\
f_1(z) &= \sum_{i=0}^{N/2}(a_i - a_{i+N/2})z^i.
\end{align}
It should be noted that these particular butterfly operations will not require ring multiplications, as their twiddle factors are all $1$.  As $n_1 = N/2$, we keep the image $f_1 = f \bmod (z^{N/2}+1)$, and use a DWT on $f_1$ to evaluate $f$ at the roots of $\Phi_1$.  We then break $f \bmod (z^{N/2}-1)$ into the images $f \bmod (z^{N/4}-1)$ and $f \bmod (z^{N/4}+1)$ in a similar fashion.  If $n_1 < N/4 < n_2$, we ignore the image $f \bmod (z^{N/4}+1)$ for the remainder of the computation.  We continue in this fashion until we have $f_s = f \bmod (z^{n_s}+1)$, after which we should have $\TFT'_{\omega,n}(f)$ in place of $f$, albeit scattered throughout our array of working space.

Performing the aforementioned modular reductions amounts to $\mathcal{O}(n)$ additions.  Accounting for the cost of the DWTs, we have that the number of multiplications becomes
\begin{equation}
\sum_{i=1}^s \tfrac{1}{2}n_i \log n_i + \mathcal{O}(n_i) < \tfrac{1}{2}n \log n + \mathcal{O}(n).
\end{equation}
A similar analysis for the ring additions due to the DWTs gives us the following complexity:
\begin{Lem}[Mateer]
$\TFT'_{\omega,n}(f)$ can be computed using $\tfrac{1}{2}n \log n + \mathcal{O}(n)$ ring multiplications plus $\mathcal{O}(n \log n)$ ring additions.
\end{Lem}
Since each of the steps of Mateer's method amounts to a series of butterfly operation, the algorithm is invertible with comparable complexity.  One advantage of Mateer's TFT is that inverting it is relatively straightforward: we invert the DWTs and then invert the butterflies.

Mateer's method, however, requires space for $N$ ring elements, as the images $f \bmod (z^{N/2}-1)$ and $f \bmod (z^{N/2}+1)$ may have maximal degree.
\section{In-place methods for computing the cyclotomic TFT}\label{sec:newTFT}

In order to compute the cyclotomic TFT in-place, it appears, unlike the Mateer TFT, that we need to use some of the information from the images $f_1, \dots, f_{i}$ towards producing the image $f_{i+1}$.  Both Sergeev's TFT and the new algorithm presented thereafter work in this manner.

Let, for $1 \leq i \leq s$,
\begin{align}
\Gamma_i(z) = \prod_{j=1}^i \Phi_i(z) \hspace{0.3cm}\text{and}\hspace{0.3cm}C_i = f \bmod \Gamma_i.
\end{align}
We call $C_i$ the {\em combined} image of $f$, as it is the result of Chinese remaindering on the images $f_1, \dots, f_i$.  We also define
\begin{align}
q_i = \left\{ \begin{array}{lll}
f && \text{ if }i=0\\
f \quo \Gamma_i && \text{ if }1 \leq i \leq s\\
\end{array}\right.
\end{align}
the quotient produced dividing $f$ by $\Gamma_i$, as well as
\begin{equation}
n_i^* = \left\{ \begin{array}{lll}
n && \text{ if } i=0,\\
n \bmod n_i && \textbf{ if } 1 \leq i \leq s,\\
\end{array}\right.
\end{equation}
Note that, as $\Phi_i \bmod \Phi_j = 2$ for $j \leq i$, we also have $\Gamma_i \bmod \Phi_j = 2^{i}$ for $j \geq i$.  Similarly, $\Gamma_i \bmod (z^K-1) = 2^i$ for $K$, a power of two at most $n_i$.

For any choice of $1 \leq i \leq s$, we have
\begin{equation}
f(z) = C_i + \Gamma_iq_i.
\end{equation}

It is straightforward to obtain $q_i$, given $f$.  Note that the degrees of any two distinct terms of $\Gamma_i$ differ by at least $n_i$, and that $\deg(q_i) < n_i$.  Thus, as $\Gamma_i$ is monic, we have that the coefficients of $q_i$ merely comprise the coefficients of the higher-degree terms of $f$.  More precisely,
\begin{equation}
q_i = \sum_{j=0}^{n_i^*-1}a_{n-n_i^*+j}z^j.
\end{equation}
By a similar argument, we also have that, for $1<i\leq s$,
\begin{equation}
q_i = q_{i-1} \quo \Phi_i.
\end{equation}
We note that $q_s = 0$.
\subsection{Computing images of $C_i$ without explicitly computing $C_i$}

We will express the combined image $C_i$, $C_i$ reduced modulo $\Phi_{i+1}$ or a polynomial $z^k-1$ with $k=2^q$, in terms of the coefficients of the images $f_1, \dots, f_i$.  To this end we introduce the following notation.  Given an integer $e$, we will let $e[i]$ refer to the $i\th$ bit of $e$, i.e.
\begin{equation}
e=\sum_{i=0}^{\lfloor \log(e) \rfloor}e[i]2^i, \hspace{0.5cm}e[i] \in \{0,1\}.
\end{equation}

Sergeev's TFT relies on the following property, albeit stated differently here than in \cite{Igor}.

\begin{Lem}\label{lem:crt2}
Fix $i$, $j$ and $k$, $1 \leq j \leq i < k \leq s$.  Suppose that $f_j=z^e$ and $f_l = 0$ for $l \neq j$, $0 \leq l \leq i$.  Let $m$ be a power of two at most $n_i$.  Then $C_{i} \bmod z^m-1$ is nonzero only if
$e[ n(l) ] = 1$ for $j < l \leq i$, in which case,
\begin{align}
C_i \bmod (z^m-1) &= 2^{i-j}z^e \bmod (z^m-1) \\
&= 2^{i-j}z^{e \bmod m}.
\end{align}
\end{Lem}

As $\Phi_k$, $k>i$, divides $z^{n_i}-1$.  Lemma \ref{lem:crt2} gives us the following corollary.

\begin{Cor}\label{lem:crt}
Fix $i$, $j$ and $k$, $1 \leq j \leq i < k \leq s$.  Suppose that $f_j=z^e$ and $f_l = 0$ for $l \neq j$, $0 \leq l \leq i$.  Then $C_{i} \bmod \Phi_{k}$ is nonzero only if
$e[ n(l) ] = 1$ for $j < l \leq i$, in which case
\begin{equation}\label{eqn:lem}
C_{i} \bmod \Phi_{k} = 2^{i-j}z^e \bmod \Phi_{k}.
\end{equation}
\end{Cor}
Given that $z^{n_k} \bmod \Phi_k = -1$, we have that
\begin{equation}\label{eqn:explicitForm}
2^{i-j}z^e \bmod \Phi_{k} = (-1)^{e[n(k)]}2^{i-j}z^{ (e \bmod n_k) },
\end{equation}
where $e \bmod n_k$ is the integer $e^*$ such that $n_k | (e-e^*)$ and $0 \leq e^* < n_k$.  The values $e[n(l)]$ can be determined from $n_l=2^{n(l)}$ and $e$ by way of a bitwise ``and'' operation.

\begin{Exam}[Example of corollary \ref{lem:crt}]
Suppose $n=86$ $(n=64+16+4+2)$, and suppose that 
\begin{align*}
f_1 = f \bmod z^{64}+1 = z^e, \hspace{0.5cm} f_2 = 0, \hspace{0.5cm} f_3 = 0,
\end{align*}
and $\deg(f) < 64+16+4$.  In this example,
\begin{equation}
C_3 = f \bmod \left[ (z^{64}+1)(z^{16}+1)(z^4+1) \right].
\end{equation}
Let $g(z) = C_3 \bmod (z^2+1)$.  Then by lemma \ref{lem:crt},
\begin{equation*}
g(z) = \left\{ \begin{array}{lll}
0 & \text{ if } e \in [0,20) \cup [24,28) \cup [32,52) \cup [56,60),\\
4 & \text{ if } e = 20,28,52, \text{ or }56,\\
4z & \text{ if } e = 21,29,53, \text{ or }57,\\
-4 & \text{ if } e = 22,30,54, \text{ or }58,\\
-4z & \text{ if } e = 23, 31, 55, \text{ or }59.
\end{array}\right.
\end{equation*}
\end{Exam}

\begin{Rem}\label{rem:nonzero}
Let $1 \leq j \leq i \leq s$.  A proportion of $2^{j-i}$ terms of $f_j$ have an exponent satisfying the nonzero criterion of lemmas \ref{lem:crt} and \ref{lem:crt2}.  Moreover, this nonzero criterion may be easily checked using a bitwise ``and'' operation.
\end{Rem}

\begin{proof}[\bf Proof of lemma \ref{lem:crt2}]
We fix $e$ and $j$ and prove by induction on $i$.

{\em Base case:} Let $i=j$.  As in the proof of lemma \ref{lem:crt} we have $C_i = 2^{1-i}\Gamma_{i-1}z^e$.  As $\Phi_l \bmod z^m-1 = 2^i$ for $m$ a power of two dividing $n_l$, it follows that
$C_i \bmod z^m-1 = z^e \bmod z^m-1$.

{\em Inductive step:}  Suppose now that the lemma holds for a fixed $i\geq j$, and consider $C_{i+1} \bmod z^m-1$, $m$ a power of two dividing $n_{i+1}$.  We suppose that $f^*_l = 0$ for $1 \leq l \neq j \leq i+1$ and $f^*_j = z^e$.  By Chinese remaindering,
\begin{equation}\label{eqmod}
C_{i+1} = C_i - \Gamma_i\left( \Gamma_i^{-1}C_i \bmod \Phi_{i+1}\right).
\end{equation}.
Reducing this expression modulo $z^m-1$ gives us
\begin{equation}
C_{i+1} = C_i - \Gamma_i\left( \Gamma_i^{-1}C_i \bmod \Phi_{i+1} \right) \bmod z^m-1.
\end{equation}

{\em Case 1:}  If $e[n(l)]=0$ for some $l$, $j<l\leq i$, then by the induction hypothesis, $C_i \bmod z^m-1 = 0$.  By lemma \ref{lem:crt}, $C_i \bmod \Phi_{i+1}=0$.  Thus we have $C_{i+1} \bmod z^m-1 = 0$ as well.

{\em Case 2:}  If $e[n(l)]=0$ for $j<l \leq i$, then by the induction hypothesis, $C_i \bmod z^m-1 = 2^{i-j}z^e \bmod z^m-1$.  By lemma \ref{lem:crt}, $C_i \bmod \Phi_{i+1} = z^e \bmod \Phi_{i+1}$
\begin{align}
&C_{i+1} \bmod z^m-1 \notag \\
&= 2^{i-j}z^e - \Gamma_i\left( \Gamma_i^{-1} z^e \bmod \Phi_{i+1} \right) \bmod z^m-1 \\
&= 2^{i-j}z^{e \bmod m} - \left( z^e \bmod \Phi_{i+1} \right) \bmod z^m-1\\
&= 2^{i-j}\left(1 + (-1)^{e[n(i+1)]} \right)z^{e \bmod m}.
\end{align}
As $1 + (-1)^e[n(i+1)]$ evaluates to $2$ if $e[n(i+1)]=1$ and $0$ otherwise, this completes the proof.
\end{proof}

Lemma \ref{lem:crt2} can be derived from lemma 1 in \cite{Igor}.

\subsection{Sergeev's in-place cyclotomic TFT}

In \cite{Igor}, Sergeev describes an arithmetic circuit for computing a cyclotomic TFT.  Such a circuit for computing $\TFT'_{\omega,n}(f)$ has a depth of $n + \mathcal{O}(1)$.  That is, it requires space for $n + \mathcal{O}(1)$ ring elements.  This algorithm, like Mateer's breaks $f$ into the images $f_i$ in linear time, and then applies a DWT on each image.  Algorithm \ref{alg:serg} describes how $f$ is broken into the images $f_i^*$.

\begin{algorithm}
\caption{Sergeev's algorithm for computing $f_1,\dots,f_s$ in place of $f$}\label{alg:serg}
\KwIn{
	$n<N$, $f(z) \in \mathcal{R}[z]$, $f=\sum_{i=0}^{n-1}a_iz^i$
}
\KwResult{ $f_1,\dots,f_s$ are written in place of $f$ }

	$(i,K) \longleftarrow (0,N)$\\

	\While{$K > n_s$}{
		{\bf Loop invariant:}  We have the images $f_j$, $1 \leq j \leq i$, and the first $n_i^*$ coefficients of $f \bmod (z^K-1)$.\\
		\If{ $K/2 = n_{i+1}$}{
			Compute $f_{i+1}$ and the first $n_{i+1}^*$ coefficients of $f \bmod (z^{K/2}-1)$ in place of the first $n_i^*$ coefficients of $f \bmod (z^K-1)$.\\
			$i \longleftarrow i+1$
		}
		\Else{
			Write the first $n_i^*$ coefficients of $f \bmod (z^{K/2}-1)$ in place of the coefficients of $f \bmod (z^K-1)$.
		}
		$K \longleftarrow K/2$
	}
\end{algorithm}

Suppose we have $f_1, \dots, f_i$, and the first $n_i^*$ coefficients of $g = f \bmod (z^K-1)$, where $K \leq n_i$ is a power of two.  Note that $K$ may be greater than $n_i^*$.  If we write
\begin{equation}
f(z) \bmod (z^K-1) = \sum_{j=0}^{K-1}b_iz^i,
\end{equation}
then
\begin{align}
f(z) \bmod (z^{K/2}-1) &= \sum_{j=0}^{K/2-1}(b_j + b_{j+K/2})z^j, \text{ and}\label{eqn:serg1}\\
f(z) \bmod (z^{K/2}+1) &= \sum_{j=0}^{K/2-1}(b_j - b_{j+K/2})z^j. \label{eqn:serg2}
\end{align}
If $K/2 > n_i^*$, Sergeev's algorithm will write the first $n_i^*$ coefficients of $f \bmod (z^{K/2}-1)$ in place of $f \bmod (z^K-1)$, i.e. we want to add $b_{i+K/2}$ to $b_i$ for $0 \leq i < n_i^*$.  Reducing $f = C_i + \Gamma_iq_i$ modulo $z^K-1$, we have
\begin{equation}
f \bmod (z^K-1) = C_i + 2^iq_i \bmod (z^K-1).
\end{equation}

\begin{Rem}
As $\deg(q_i) < n_i^* < K/2$, the coefficients $b_{j+K/2}$, $0 \leq j < n_j^*$ depends strictly on $C_i \bmod (z^K-1)$.
\end{Rem}
We use lemma \ref{lem:crt2} to determine the {\em contribution} of each coefficient of $f_j$, $1 \leq j \leq i$, towards $b_{i+K/2}$.  Namely, if a term $cz^e$ appearing in $f_j$ satisfies the nonzero criterion of lemma \ref{lem:crt2}, then it will have a contribution of $2^{i-j}c$ towards $b_{i+K/2}$.  For $0 \leq i < n_i^*$, we compute the sum of these contributions towards $b_{i+K/2}$ and then add that to $b_i$.  In order to minimize the number of multiplications necessary in order to compute the weights $2^{i-j}$, we sum the contributions from $f_1$, then multiply that sum by $2$, then add to that all the contributions from $f_2$, again multiply that by 2, and so forth.

If instead, $K \leq n_i^*$, we have that $K$ must equal $n_{i+1}$, we write $f_{i+1}$ and the first $n_{i+1}^*$ coefficients of $f \bmod (z^{K/2}-1)$ in place of $b_j$, $0 \leq j < n_i^*$.  For $0 \leq j < n_{i+1}^*$, we have both $b_i$ and $b_{i+K/2}$, and can compute the $z^i$ coefficient of $f_{i+1}$ and $f \bmod (z^{K/2}-1)$ in their places accordingly.  What remains is to compute the last $n_i - n_i^*$ coefficients of $f_{i+1}$.  This amounts to computing $b_{i+K/2}$ by way of lemma \ref{lem:crt2}, as in the first case.

\section{A new in-place cyclotomic TFT algorithm}\label{sec:ctft}

In order to compute a new coefficient of an image of $f$ in Sergeev's algorithm, one has to effectively make a pass through the array in order to sum the contributions from images $f_j$.  Instead of summing the contributions and then adding that into our array, we will add contributions directly back into our array of coefficients.  In order to do this, we will work with the {\em weighted} images,
\begin{equation}
f_i^*(z) = 2^{-i+1}f_i,
\end{equation}
this allows us to add all the contributions to a new polynomial image with a single pass over our array.  Moreover, unlike Sergeev's algorithm, we will forego producing part of the images $f \bmod (z^K-1)$, $K$, a power of two.  We will compute $f_{i+1}$ immediately after producing $f_i$.

Our transform has three steps: we first iteratively compute the remainders produced by dividing $q_{i-1}$ by $\Phi_i$,
\begin{equation}
r_i = q_{i-1} \bmod \Phi_i, \hspace{0.3cm} 1 \leq i \leq s.
\end{equation}
in place of $f$; we then iteratively write $f_i^*$ in place of $r_i$ for $i=1,2,\dots,s$; lastly we reweigh the weighted images $f_i^*$ to get $f_i$ and compute a DWT of each image $f_i$ separately to give us the weighted evaluation points.

%
%

\subsection{Breaking $f$ into the remainders $r_1, \dots, r_s$}\label{sec:ri}

We first break $f=q_0$ into its quotient and remainder dividing by $z^{n_1}+1$,
\begin{align}
r_1 = f &\bmod \left(z^{n_1}+1\right) = \sum_{i=0}^{n_1-1}\left( a_i - a_{i+n_1} \right)z^i,\\
q_1 = f &\quo \left(z^{n_1}+1\right) = \sum_{i=0}^{n_1^*}a_{i+n_1}z^i,
\end{align}
where $a_i=0$ for $i \geq n$.  This can be done in place with $n_1^*$ subtractions in $\mathcal{R}$.  We then similarly break $q_1$ into $r_2$ and $q_2$, then $q_2$ into $r_3$ and $q_3$, and continue until we have $r_1, \dots, r_{s-1}$ and $q_{s-1}$.  Since $\deg(q_{s-1})<n_s = \deg(\Phi_s)$, $r_s$ is exactly $q_{s-1}$.


\subsection{Writing the weighted images $f_i^*$ in place of the remainders $r_i$}

We first note that $f_1^*$ is precisely $r_1$.  We will iteratively produce the remaining weighted images.

Suppose, at the start of the $i\th$ iteration, we have $f_1^*, \dots, f_{i}^*$, and $r_{i+1}, \dots, r_s$, and we want to write $f_{i+1}^*$ in place of $r_{i+1}$.  We have
\begin{align}
f_{i+1}^* &= 2^{-i}f \bmod \Phi_{i+1}, \\
&=2^{-i}\left(\Gamma_{i}q_i + C_{i} \right) \bmod \Phi_{i+1}, \\
&= \left( q_i + 2^{-i}C_{i} \right) \bmod \Phi_{i+1},\\
&= r_{i+1} + \left( 2^{-i}C_i \bmod \Phi_{i+1}. \right)\label{eqn:ith_iteration}
\end{align}

Unfortunately, we don't have the combined image $C_{i}$, but rather the weighted images $f_j^*$, $1 \leq j \leq i$, from which we can reconstruct $C_i$.  We would like to be able to compute $C_i \bmod \Phi_{i+1}$ in place from the weighted images $f_j^*$.

Corollary \ref{lem:crt} tells us the {\em contribution} of $f^*_i$ towards subsequent images.  If $e$ satisfies the non-zero criterion of the lemma, then by \eqref{eqn:ith_iteration}, a term $c_{j,e}z^e$ of $f^*_j$, $j\leq i$ will contribute $\tfrac{1}{2}(-1)^{e[n(i+1)]}z^{ (e \bmod n_{i+1}) }$ to $f^*_{i+1}$.  In order to make the contributions have weight $\pm 1$, we instead first reweigh $r_i$ by $2$ and compute $2f^*_i$, and then divide by $2$ thereafter.  {\tt\ref{proc:addContribution}} adds the contribution of $f^*_1,\dots, f^*_{i-1}$ to $f^*_i$.
\begin{procedure}
\caption{AddContributions($\textbf{a}$, $n$, $i$ )}\label{proc:addContribution}

\KwIn{
\begin{itemize}[noitemsep,nolistsep]
\item $n = \sum_{j=1}^s n_j$.
\item ${\bf a}$, a length-$n$ array containing $f^*_1, \dots, f^*_{i-1}$ and $2r_i$, in that order.
\end{itemize}
}
\KwResult{
	The contribution of $f^*_1, \dots, f^*_{i-1}$ towards $2f^*_i$ are added to $2r_i$.  As a result we will have $2f^*_i$ in place of $2r_i$.
}

$\delta_{\tt out} \longleftarrow \sum_{j=1}^{i-1}n_j$

\For{$j \longleftarrow 1$ \KwTo $i-1$}{
	\tcp{Add contribution of $f^*_j$ to $f^*_i$}
	$\delta_{\tt in} \longleftarrow \sum_{k=1}^{j-1}n_j$\\
	\For{$e \longleftarrow 0$ \KwTo $n_j-1$}{
		\If{$e[n(l)]=0$ \text{ for } $j < l < i$}{
			$\a( \delta_{\tt out} + (e \bmod n_i))$ {\tt +=} $(-1)^{e[n(i)]}\a( \delta_{\tt in} + e)$
		}
	}
}
\end{procedure}

According to corollary \ref{lem:crt}, $f^*_j$, $0 \leq j \leq i$, only a proportion of $2^{i-j}$ of the terms of $f^*_j$ will have a non-zero contribution to $f^*_{i+1}$.  Thus the total cost of adding contributions of $f^*_j$ towards $f^*_i$, $i>j$, is less than $2\#f^*_j = 2n_j$.  It follows that the total additions and subtractions in $\mathcal{R}$ required to add all these contributions is bounded by $2n$.  Since {\tt\ref{proc:addContribution}} only scales array ring elements by $\pm 1$, we have the following complexity:

\begin{Lem}\label{ref:contrib}
Calling {\tt\ref{proc:addContribution}}$({\bf a}, n, i)$ for $i$ from 1 to $s$ entails no more than $2n$ ring additions and no ring multiplications.
\end{Lem}

In the manner we have chosen to add these contributions, we will have to make $s-1$ passes through our array to add them all.  One way we could avoid this is to instead add all the contributions from $f^*_1$, and then all the contributions from $f^*_2$, and so forth, adding up all the contributions from a single term at once.  We could use that a term $c_{j,e}z^e$ of $f^*_j$ that does not contribute towards $f^*_{i+1}$ will not contribute to $f^*_k$ for any $k > i$.  This would reduce the number of passes we make through the larger portion of the array, though the cache performance of potentially writing to $s-1$ images at once raises questions.

When adding contributions to $f^*_i$, any two terms $c_{j,e}z^e$ and $c_{j,e^*}z^{e^*}$ of $f^*_j$ whose bits $e[n(l)], e^*[n(l)]$ agree for $j < l < i$ will both contribute to $f^*_i$ in the same fashion (i.e. we will either add or subtract both coefficients $c_{i,e}$ and $c_{i,e*}$ to an image, or do nothing).  Thus we need only inspect the non-zero criterion of one exponent $e$ in a block of exponents $kn_{i-1} \leq e < (k+1)n_{i-1}$.  Similarly, we need only inspect one exponent in a contiguous block of $n_i$ exponents in order to determine their shared value of $(-1)^{e[n(i)]}$.

Given an exponent that does not satisfy the non-zero criterion, our implementation will generate the next exponent that does satisfy the non-zero criterion, by way of bit operations.

\begin{procedure}
\caption{BreakIntoImages($\textbf{a}$, $n$): An in-place algorithm to compute a vector of images of $f$}\label{proc:break}\DontPrintSemicolon
\KwIn{
 ${\bf a}$, a length-$n$ array containing the coefficients of $f=\sum_{i=0}^{n-1}a_iz^i$, where $n=\sum_{i=1}^s n_i$.s
}
\KwResult{
The images $f_i$, $1 \leq i \leq s$, are written in place of $f$.
}\vspace{0.3cm}

\tcp{Write $r_1, \dots, r_s$ in place of $f$}
$m \longleftarrow 0$\\
\For{$i \longleftarrow 1$ \KwTo $s$}{
	\For{$j \longleftarrow m$ \KwTo $n-n_i-1$}{
		${\bf a}(j) \longleftarrow {\bf a}(j)-{\bf a}(j+n_i)$
	}
	$m \longleftarrow m+n_i$
}\vspace{0.3cm}

\tcp{Write the weighted image $f^*_i$ in place of $r_i$, for $1 \leq i \leq s$}
$m \longleftarrow 0$\\
\For{$i \longleftarrow 2$ \KwTo $s$}{
	\lFor{$j \longleftarrow m$ \KwTo $m+n_i-1$}{ ${\bf a}(j) \longleftarrow 2{\bf a}(j)$ }\\
	{\tt\ref{proc:addContribution}}$({\bf a}, n, i, {\tt true})$\\
	\lFor{$j \longleftarrow m$ \KwTo $m+n_i-1$}{ ${\bf a}(j) \longleftarrow \tfrac{1}{2}{\bf a}(j)$ }\\
	$m \longleftarrow m + n_i$
}\vspace{0.3cm}

\tcp{Reweigh $f^*_i$ to get $f_i$}
$m \longleftarrow n_1$\\
\For{$i \longleftarrow 2$ \KwTo $s$}{
	\lFor{$j \longleftarrow m$ \KwTo $n-1$}{ ${\bf a}(j) \longleftarrow 2{\bf a}(j)$ }\\
	$m \longleftarrow m+n_i$
}

\end{procedure}

Procedure {\tt\ref{proc:break}}$(\a, n)$ breaks $f$ into the images $f_i$.  The procedure effectively has three sections.  In the first section of the algorithm, we break $f$ into the remainders $r_i$.  Producing $r_i$ entails $n_i^* < n_i$ additions, and so producing all the $r_i$ entails less than $\sum_{i=1}^s n_i = n$ ring additions.

In the second section we write the weighted images $f_i^*$ in place of the $r_i$.  Adding all the contributions, per lemma \ref{ref:contrib}, requires $2n$ additions.  We reweigh the last $n-n_1$ coefficients of $f$ by $2$, then by $\tfrac{1}{2}$.  This constitutes less than $n$ such multiplications.

In the third section we reweigh the weighted images $f_i^*$ to get the images $f_i$.  This entails less than $n$ multiplications by $2$.  This gives us the following complexity:

\begin{Lem}\label{lem:break}
Procedure {\tt\ref{proc:break}}$(\a, n)$ entails no more than $3n$ additions and $2n$ multiplications by $2^{\pm 1}$.
\end{Lem}

Thus, like Mateer's and Sergeev's algorithms, the TFT algorithm presented amounts to linearly many ring operations plus the cost of performing FFTs of size $n_i$, $1 \leq i \leq s$.
Inverting the TFT is straightforward.  Given the images $f_1, \dots, f_s$, we can reobtain the polynomial $f$ by effectively reversing the steps of procedure {\tt\ref{proc:break}}.  Since {\tt\ref{proc:break}} strictly entails 1) multiplying arrays elements by nonzero scalars and 2) adding or subtracting one array element from another, it is straightforward to undo each of the steps to get back $f$, with similar complexity.

\section{Computing the bit-reversed TFT by way of the cyclotomic TFT}\label{sec:affine}

The bit-reversed TFT has the property that, for $m<n$, $\TFT_\omega^m(f)$ is merely the first $m$ entries of $\TFT_\omega^n(f)$, whereas the cyclotomic TFT does not, in general.  Hence the bit-reversed TFT lends itself more readily to multivariate polynomial arithmetic.  We show how an algorithm for computing the cyclotomic TFT may be modified to compute a bit-reversed TFT.

As before, let $N=2^p>n=\sum_{i=1}^s n_i = \sum_{i=1}^s 2^{n(i)}$, $\omega \in \mathcal{R}$ be an $N\th$ principal root of unity, and let $\omega_i = \omega^{N/2n_i}$ be a root of $\Phi_i$.  Define
\begin{equation}
\Omega_i = \prod_{l=1}^{i}\omega_l \hspace{0.5cm}\text{ and }\hspace{0.5cm} \Psi_j(z) = z^{n_j} - \Omega_{j-1}^{n_j}
\end{equation}
for $0 \leq i \leq s$, $0 < j \leq s$.  $\TFT_{\omega,n}(f)$ is comprised of the evaluation of $f(z)$ at the roots of the polynomials $\Psi_j(z)$.  If $n_1 + \dots + n_{j-1} \leq l < n_1 + \dots + n_{j}$, we have that $\omega^{[l]_p}$ is a root of $\Psi_j$.  To see this, write $l = n_1 + \dots + n_{j-1} + l'$, $0 \leq l' < n_j$, and observe that
\begin{align}
\omega^{[l]_p n_{j}} &= \omega^{(N/n_1 + \dots + N/n_{j-1})n_j}\omega^{[ l ]_p n_j} \\
&= (\omega_1 \cdots \omega_j)^{n_j} \omega^{[l]_p n_j},\\
&= \Omega_{j-1}^{n_j}\omega^{[l]_p n_j} \label{eqn:OMEGA}
\end{align}
As $l'<n_j$, $\omega^{[l']_p}$ has order dividing $n_j$, and so \eqref{eqn:OMEGA} is precisely $(\Omega_{j-1})^{n_j}$.  Every $n_j-th$ root of unity in $\mathcal{R}$ is of the form $\omega^{[k]_p}$, where $0 \leq k < n_j$.  In particular $\omega^{[l']_p}$ is an $n_j-th$ root of unity.  Thus $\omega^[l]_p\omega^{[l']_p}$ is a root of $\Psi_j$.  As the binary representations of $n_1 + \dots + n_{j-1}$ and $l'$ do not share any $1$ digits at the same precision,
\begin{equation}
[n_1 + \dots + n_{j-1}]_p + [l']_p = [l]_p,
\end{equation}
and so $\omega^{[l]_p}$ is a root of $\Psi_j$.

Consider the affine transformation $z \mapsto \Omega_{s}z$.  Then
\begin{align}
\Psi_i( \Omega_s z) &= \Omega_s^{n_i}\left(z^{n_i} - \Omega_{i-1}^{n_i}\right) \\
 & = \Omega_{i-1}^{n_i}\left( \prod_{j=i}^s \omega_j^{n_i} \right)\left( z^{n_i} - \Omega_{i-1}^{n_i} \right)\\
 & = -\Omega_{i-1}^{n_i}\left( z^{n_i} + 1\right) = -\Omega_{i-1}^{n_i}\Phi_i.
\end{align}

Thus, for a polynomial $f(z)$, $f(z) \bmod \Phi_i(z) = f(z) \bmod \Psi_i( \Omega_s z)$.  We can break $f(z)$ into its images modulo the polynomials $\Phi_i$ as follows:
\begin{enumerate}
\item Replace $f(z)$ with $f( \Omega_s z)$.
\item Break $f( \Omega_s z)$ into its images modulo $\Phi_i(z)$ per a cyclotomic TFT method.  Equivalently, this gives us the images $f( \Omega_sz ) \bmod \Psi_i( \Omega_s z)$.
\item Apply the transformation $z \mapsto \Omega_s^{-1}z$ to each image to give us $f(z) \bmod \Psi_i(z)$ in place of $f( \Omega_s z) \bmod \Psi_i( \Omega_s z)$.
\end{enumerate}

Note that the affine transformations of steps 1 and 3 both have complexity $\mathcal{O}(n)$.  We can get $\TFT_{\omega,n}(f)$ from the images $f \bmod \Phi_i$ by applying a DWT to each image $f \bmod \Phi_i$.  As each step here is invertible, we can invert a bit-reversed TFT using an inverse cyclotomic TFT algorithm.  We can similarly invert a bit-reversed TFT using an inverse cyclotomic TFT algorithm.

\section{Acknowledgments}
Thanks to: Igor Sergeev for bringing \cite{Todd} and \cite{Igor} to my attention, and for his assistance in helping me understand his TFT algorithm; Mark Giesbrecht for feedback; Joris van der Hoeven for posing the question of non-invertible TFTs; and Dan Roche for providing his implementation of bit-reversed TFT algorithms.

\bibliographystyle{abbrv}
\bibliography{tft_issac}  



\end{document}